 \documentclass[prd,a4paper,
twocolumn,amsmath, amssymb, floatfix,
nofootinbib,wrapfloat byrevtex]{revtex4}

 \usepackage{amsmath,mathrsfs,upgreek,graphicx,placeins,float}
 \usepackage{color}
 \usepackage{booktabs}
 \usepackage{multirow}

 \newcommand{\beq}[1]{\begin{equation}\label{#1}}
 \newcommand{\eeq}{\end{equation}}
 \newcommand{\bea}[1]{\begin{eqnarray}\label{#1}}
 \newcommand{\eea}{\end{eqnarray}}
  \newcommand{\bec}[1]{\begin{center}\label{#1}}
 \newcommand{\eec}{\end{center}}
 \newcommand{\mm}{{\rule[1.5pt]{3pt}{0.6pt}}}

\begin{document}

 \title{Gravity Induced Spontaneous Radiation}
 \author{Ding-fang Zeng}
 \email{dfzeng@bjut.edu.cn}
 \affiliation{Beijing University of Technology, Chaoyang, Bejing 100124, P.R. China,
 \\
 Niels Bohr Institute, Blegdamsvej 17, Copenhagen, 2300, Denmark}
 \begin{abstract}
We suggest that behind the black hole information paradox is a new and universal radiation mechanism, Gravity Induced Spontaneous Radiation, or GISR hereafter. This mechanism happens to all kinds of compositional objects and it requires only their microscopic structure as the basis. It's always accompanied with such inner structures' variation and allows for explicitly hermitian hamiltonian description. For black holes, by Wigner-Wiesskopf approximation we show that such a radiation has a thermal spectrum exactly the same as hawking radiation; while through numeric integration, we show that the variation of the radiation particles' entropy exhibits all features of Page curve as expected. We also provide exact and analytic solutions to the Einstein equation describing microscopic structures required by the GISR of black holes and show that, after quantisation the degeneracy of wave functions corresponding with those solutions are consistent with the area law formula of Bekenstein-Hawking entropy.
\end{abstract}
 \maketitle
 
 Among his many works, the discovery of black hole radiation and the related information missing puzzle may be S. Hawking's most profound contribution to modern theoretical physics \cite{hawking1975cmp}. Historically, calculations of the gray-body factor of black-brane radiation provide the most hard-core evidence for Anti-de Sitter/Conformal Field Theory correspondence \cite{AdScftReview99}. Recently, explorations of the information missing puzzle bring us new ideas such as ER=EPR\cite{EREPR} and Island formulae et al \cite{ReplicaWormholeReview2006}. This work will, additionally, propose that underlying this radiation and the information missing puzzle is a universal mechanism rarely investigated in the literature and encodes rich information on black hole microscopic structure, singularity theorem and quantum gravity. To understand this statement better, we need going back to S. Mathur's small correction theorem \cite{fuzzball2009mathur} for a while. 
 
\begin{figure}\begin{center}
\includegraphics[totalheight=39mm]{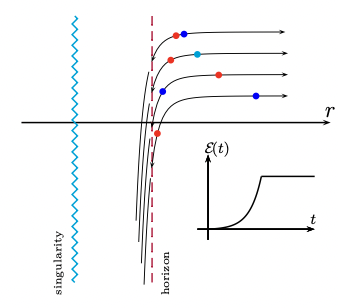}~~
\includegraphics[totalheight=39mm]{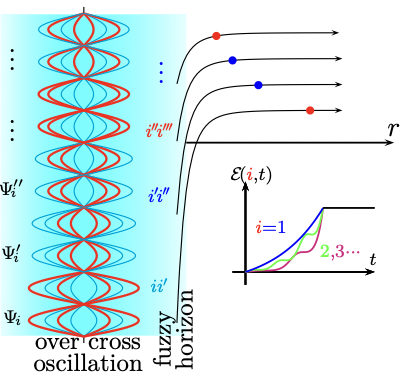}
\end{center}
\caption{In conventional understanding, hawking particles arise from vacuum fluctuation and escape from the no-hair horizon. Collecting and measuring their energy provide us energy-time curves universal to all microscopic black holes.  In GISR, radiations arise from their coupling with the microscopic state of black holes embodied in their Bekenstein-Hawking entropy. Collecting and measuring their energy will provide us a microscopic state dependent energy-time curve. In this work, microscopic state of the black hole exhibits as many modes of over cross oscillation of the black hole matter contents measured in proper time.}
\label{figMechansimComparison}
\end{figure}

 Mathur's theorem says that as long as Hawking radiation happens through pair production and escaping mechanism around the no-hair horizon, then no matter how ingeniously small corrections are added to the evolution of a black hole, the entropy of its hawking particles will increase monotonically until the black hole evaporates away. Basing on this theorem, it's very natural to conclude that some type of new radiation mechanism is the only way to solve the information missing puzzle \cite{dfzeng2018a,dfzeng2018b,stojkovic2020unp}. We will call this mechanism GISR, as dubbed in the abstract of this work, see FIG.\ref{figMechansimComparison} for references.  
We conjecture that GISR is a universal radiation mechanism that happens to all compositional objects instead of black holes exclusively, see the appendix for reasons for this mechanism's universality. However, in this work black holes will be taken as proxies for all such objects and we will talk about mainly three aspects of this mechanism. That is, its explicitly hermitian hamiltonian description, its resolution to the information missing puzzle and its implication for black hole inner structures and quantum gravitation theory. 

Just as many other no-go theorems forbidding some working directions, e.g. Coleman-Mandula theorem's forbidding boson-fermion symmetries in classic non-graded Lie algebra \cite{WeinbergSUSYtxtbook}, Mathur's theorem also contains implicit assumptions, especially, the factorisation doing which writes the intermediate state of an evaporating black hole system as direct product of the black hole's and that of the hawking particles. This is a very important assumption adopted implicitly in many arguments for information missing puzzle, such as AMPS(S) firewall argument \cite{fireworksAMPS2012,fireworksAMPS2013}. Just as we pointed out in refs.\cite{dfzeng2021} and will emphasise in the current work, when this assumption is get rid of and the intermediate state is written as entanglement superposition of B\&R (black hole and radiation) combinations of different mass ratio, the missing information will be retrieved very naturally. Refs.\cite{Papadodimas2013,Stephen1302,Stephen1308,Stephen2022,NVW2013a,NVW2013b} point out that Mathur's theorem contains another kind of factorisation which writes $U\sum_d\psi_d(i)\rightarrow\sum_{d'}U\psi_{d'}(i)$, where $U$ means evolution and $d$ distinguishes macroscopic states of the radiating black hole due to the back reaction from hawking particles. These works argue that unitarity of the full quantum state's evolution does not imply that of the single quantum state basis. Getting rid of this assumption has the potential to retrieve information from hawking particles. 

Many technique details referred to in this work can be found in our long paper \cite{dfzeng2021}, new points of this work is, i) for the first time we realise and express explicitly that, GISR may be a universal mechanism which happens to all kinds of compositional objects. Black holes are just a special example; ii) for the first time, we realise and write out properly the concrete form of couplings between the radiating black hole and radiated particles through GISR. Our previous writing 
\beq{}
H_\mathrm{int} =\frac{\mathrm{SimilarityFactor}[\psi_\mathrm{ini},\psi_\mathrm{fnl}]}{\sqrt{G}}*b^\dagger_{uv}a_{u-v}
\eeq
has the fault of all black holes radiating at a uniform rate. In this work, we change the couplings into  
\beq{}
H_\mathrm{int}=\frac{\mathrm{SimilarityFactor}[\psi_\mathrm{ini},\psi_\mathrm{fnl}]}{G*\mathrm{Max}[M_\mathrm{ini},M_\mathrm{fnl}])}*b^\dagger_{uv}a_{u-v}
\eeq 
which will allow the black holes to radiate at a rate ${\propto}M^{-2}$, thus coincide better with the black hole thermodynamics and Stefan-Boltzmann law, $d^2M/dt*dA=\sigma*T^4$. This coupling has intuitive explanation that GISR happens like a classic diffusion. The bigger the radiating body is, the slower it diffuses; iii) only in the current work, we realise and express definitely what was done erroneously in the conventional argument for information missing in Hawking radiation is kind of Schr\"odinger’s cat misunderstanding.

\section{\bfseries Explicitly Hermitian Hamiltonian Description} Different from the usual atomic radiation which happens through electric dipole couplings between the atom and photons,  GISR happens through gravitational monopole couplings between the black hole and to-be radiated particles, 
\beq{}
H=H_\mathrm{BH}+H_\mathrm{vac}+H_\mathrm{int}
\label{HamiltonianA}
\eeq
\beq{}
=\begin{pmatrix}w^i\\&w_{\mm}^{j}\\&&\ddots\\&&&{\scriptstyle\it0}^{\scriptscriptstyle\it1}\end{pmatrix}\!+\!\sum_q\hbar\omega_qa^\dagger_qa_q
{+}\!\!\sum_{u\;\!\!^n\mm v^\ell}^{\hbar\omega_q=}g_{u\;\!\!^n v^\ell}b^\dagger_{u\;\!\!^n v^\ell}a_q
\label{HamiltonianB}
\eeq
\beq{}
g_{u\;\!\!^nv^\ell}{\propto}{-}\frac{\hbar}{G\{M_u,M_v\}^\mathrm{max}}\mathrm{Siml}\{\Psi[M_{u\;\!\!^n}\!(r)],\Psi[M_{v\;\!\!^\ell}\!(r)]\}
\label{monopoleCouplingStrength}
\eeq 
In this hamiltonian, $w$, $w_\mm{=}w{\mm}{\scriptstyle1}$ et al denote the degeneracy or eigenvalue of black hole mass/energy regarding contexts; $i$, $j$ et al distinguish microscopic states of equal mass black holes, $i=1,2\cdots,w$, $w=\exp[\frac{4\pi r_{\!h}^2}{4G\!_N}]$, $j=1,2\cdots,w{\mm}{\scriptstyle1}, w{\mm}1=\exp[\frac{4\pi r_{\!h}^{\prime2}}{4G\!_N}]$; the symbol ${\scriptstyle\it0}^{\scriptscriptstyle\it1}$ represents quantum state of the system when the black hole totally evaporate away; $a_q^\dagger$ \& $a_q$ are operators describing the vacuum fluctuation, $b^\dagger_{u\;\!\!^nv^\ell}$ \& $a_q$ take responses for the black hole mass/energy level's lowering or increasing and the vacuum fluctuation mode $\omega^n_q$'s realisation or inverse.  The analogue logic behind the coupling $g_{uv}$ and the electric dipole coupling is $\frac{\hbar}{GM}\mathrm{Siml}\{\cdots\}\sim{}\vec{d}{\cdot}\vec{A}$. The similarity factor plays similar roles of electric-dipoles while $\frac{\hbar}{GM}$ measures the energy reward if the black hole radiates a particle away or absorbs one in. More massive the black hole is, more negligible it will be rewarded from such a process. See ref\cite{dfzeng2021} for the definition of similarity factors in our concrete model of black hole inner structures.  In some sense, GISR is a diffusion-like phenomena, so we conjecture it a universal effects. For the same reason, the energy profit becomes zero when the black hole becomes infinitely massive or its horizon covers the the whole universe. Well as is motivated physically, we must note here that the concrete form of \eqref{monopoleCouplingStrength} is an intuitive ansatz. Currently we still do not know if there is any first principle from which one can derive it out exactly. The idea of reference \cite{Stephen2022} provides a possible working direction.

\begin{figure}[h]\begin{center}
\includegraphics[totalheight=25mm]{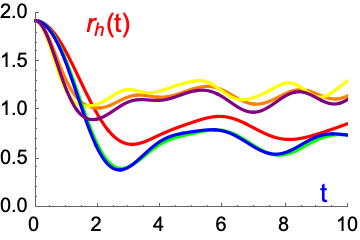}
\includegraphics[totalheight=25mm]{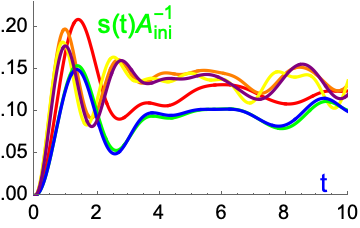}
\end{center}
\caption{Evolutions of the horizon size and radiation products' entropy of 6 degenerating black hole eigenstate following from exact numeric integrations of eq\eqref{SchrodingerEq} with initial condition like \eqref{initialCondition}. $A_\mathrm{ini}$ is the horizon area of initial black holes.}
\label{figWsixRadiationExample}
\end{figure}

Focusing on spherically symmetric radiations only, so the hawking particles' spatial-momentum can be ignored and their quantum state will be characterised by their energy exclusively\footnote{As was firstly noted by D.Page and emphasised in \cite{Stephen1302,Stephen1308}, during most hawking radiation events, the center of mass position and velocity of the BH will change unavoidably due to radiation back reaction. Considering such effect helps us little in understanding physics involved in the information missing puzzle. However, such effect will change indeed some details of our work, e.g. it will make Rabi-like oscillations on our $r_h(t)$ curve and $s(t)$ curve in FIG.\ref{figWsixRadiationExample} almost not happen.}, the basis of Hilbert space for an evaporating black hole and the corresponding radiations can be written as
\begin{align}
\{w^i\otimes{\it\phi},w^j_{\mm}\otimes\omega_1^i,w^k_{{\mm\,\mm}}\otimes\omega_1^j\omega_1^i,w^j_{{\mm\,\mm}}\otimes\omega_2^i,\cdots
\nonumber\\
,u\;\!\!^n\otimes q^k{p}^j\cdots{o}^i(u{+}q{+}p\cdots{+}o=w),\cdots\;
\\
{\scriptstyle\it0}^{\scriptscriptstyle\it1}\otimes\omega_1^z\cdots\omega_1^j\omega_1^i,{\scriptstyle0}^{\scriptscriptstyle1}\otimes\omega_1^y\cdots\omega_2^i,\cdots,{\scriptstyle0}^{\scriptscriptstyle1}\otimes\omega_w^i
\}\nonumber
\end{align}
On this basis, the quantum state of a black hole and its radiations at arbitrary middle epoch can be written as
\beq{}
|\psi(t)\rangle=\sum_{u=w}^0\sum_{n=1}^{u}\sum_{{\scriptscriptstyle\sum}o^i}^{w{\mm}u}
e^{-iut-i\omega t}c_{u\;\!\!^n}^{\vec{\omega}}(t)|u\;\!\!^n\otimes\vec{\omega}\rangle
\label{WaveFunctionRadiation}
\eeq
where ${\vec{\omega}}\equiv\{o^i,\cdots p^jq^k\}$ is an abbreviation of the radiation particles' quantum state, with the index $i,j$ and $k$ et al inherited from the parent body $w^i$, $w^j_{\mm}$ and $v^k$ et al correspondingly and the total energy given by $\omega{\equiv}o+{\cdots}p+q=w-u$. Evolutions of this wave function are determined by the standard Schr\"odinger equation as follows
\beq{}
i{\partial}_tc^{\vec{\omega}}_{u\;\!\!^n}(t){=}
\sum_{v\neq u}^{v+{\scriptscriptstyle'\!}\omega=w}\sum_{\ell=1}^{v}g_{u\;\!\!^n v^\ell}c^{{\scriptscriptstyle'\!}{\vec{\omega}}}_{v^\ell}(t)
\label{SchrodingerEq}
\eeq
where $\hbar$ has been set to $1$ and ${{\scriptscriptstyle'}{\vec{\omega}}}$ differs from $\vec{\omega}$ only by the last emitted or absorbed particles. Without loss of generality, we will set
\beq{}
c_{w^1}^{\scriptscriptstyle\it\phi}(0)=1,c_{w^{i\neq1}}^{\scriptscriptstyle\it\phi}(0)=0,c_{u\;\!\!^n}^{\vec{\omega}\scriptscriptstyle\neq\phi}(0)=0
\label{initialCondition}
\eeq
That is, we let our black hole lie on eigenstate $w^1$ at initial time $t=0$. 

For the first one or few particles' radiation, we can set all $c_{v\;\!\!^\ell}^{{\scriptscriptstyle'\!}\vec{\omega}\neq\phi,o{\scriptscriptstyle1}}{=}0$ and focus on the evolution of $c^\phi_{w^1}(t)$ and $c_{u\;\!\!^n}^{o{\scriptscriptstyle1}}(t)$ exclusively. In this case, by the standard Wigner-Wiesskopf approximation \cite{dfzeng2021,ScullyQOtextbook}, we can easily prove that
\beq{}
c^\phi_{w^1}(t){=}e^{-\Gamma t},
c_{u\;\!\!^n}^{o{\scriptscriptstyle1}}(t){=}\frac{ig_{u\;\!\!^n\!w\;\!\!^1}}{\Gamma}(e^{\mm\Gamma t}\mm1)
,c_{v\;\!\!^\ell}^{{\scriptscriptstyle'\!}\vec{\omega}\neq\phi,o{\scriptscriptstyle1}}{=}0
\label{WignerWiesskopfC}
\eeq
\beq{}
\sum_{n}\frac{g^{2}_{u\;\!\!^n\!w\;\!\!^1}}{\Gamma^2}\approx\frac{u[{=}e^{4\pi G_{\scriptscriptstyle N}(M-o{\scriptscriptstyle1})^2}]}{w_\mm({=}e^{4\pi G_{\scriptscriptstyle N}M^2}\mm1){+}\cdots2{+}1}\approx\frac{e^{-8\pi GMo{\scriptscriptstyle1}}}{e^{4\pi G_{\scriptscriptstyle N}M^2}}
\label{guwOgammasq}
\eeq
where $\Gamma{\equiv}\!\sum_{u,n}^{u{\!\neq\!}w}|g_{w\;\!\!^1\!u\;\!\!^n}\!|^2\delta(w{-}u{-}o{\scriptstyle1})$ and the first step in \eqref{guwOgammasq} uses the fact that $g_{u\;\!\!^n\!w\;\!\!^1}$ is approximately equal for most radiation channels. Setting $8\pi G_{\scriptscriptstyle N}M\equiv(k_{\scriptscriptstyle\!B\!}T)^{-1}$, the energy of the radiated particle reads
\beq{}
\langle E\rangle_{t\rightarrow}^{\infty}{=}\!\!\!\!\!\sum_{o{\scriptscriptstyle1},n}^{u+o{\scriptscriptstyle1}=w}
\!\!\!{o\scriptstyle1}|c_{u\;\!\!^n}^{o{\scriptscriptstyle1}}|^2
{=}\!\sum_{k}\!\frac{k\omega e^{-\frac{k\omega}{k_{\scriptscriptstyle\!B\!}T}}}{e^{{\mm}\frac{k\omega}{k_{\scriptscriptstyle\!B\!}T}}{+}\cdots1}{=}\frac{\omega}{e^{\frac{k\omega}{k_{\scriptscriptstyle\!B\!}T}}{\pm}1}
\label{powerSpectrum}
\eeq
The second step assumes that the radiated particle is quantised so that $o{\scriptstyle1}=k\omega$ with $k=0,1$ for fermions and $k=0,1,2\cdots$ for bosons. Eq\eqref{powerSpectrum} implies that the power spectrum of GISR of black holes is completely the same as Hawking radiation.  By our conjecture GISR happens to all kinds of compositional objects, whose effective temperature can be read out from the derivations in \eqref{guwOgammasq}
\beq{}
k_\mathrm{B}T_\mathrm{GISR}=W\frac{dM}{dW}, k_{\scriptscriptstyle\!B}\log W=S.
\eeq
For black holes, this implies $k_\mathrm{B}T_\mathrm{GISR}=\frac{1}{8\pi GM}=k_\mathrm{B}T_\mathrm{hwk}$ very naturally. While for non black hole objects, to get this temperature, we need to know the radiating bodies mass v.s. entropy as a prerequisite. It is worth emphasising that the thermal feature of GISR of black holes is a universal short-time behaviour of the radiating master's exponential degeneracy of microscopic states instead of the Wigner-Wiesskopf approximation we adopted. The exactly exponential degeneracy of the black holes' microscopic state is due to their big mass value and owning of horizon instead of any artificial assumption. The latter fact allows their microscopic state wave function be written as the direct product of many concentric shells', all are constrained to peaks inside the horizon due to conditions eq\eqref{eomShellClassic} latter. Different from black holes, eigenstates of most other compact objects degenerate only approximately, usually with exponentially small mass/energy split. However, this split does not form obstacles for us to describe their GISR with a uniform hamiltonian of eqs\eqref{HamiltonianA}-\eqref{monopoleCouplingStrength}.

For the long term behaviour of the black hole and its GISR, we only need to integrate equations \eqref{SchrodingerEq}-\eqref{initialCondition} numerically. We provide in FIG.\ref{figWsixRadiationExample} results of this integration explicitly. From the figure, we easily see that the variation of the radiation particles' entropy has indeed first increasing then decreasing feature, just as Page curve manifests for unitarily evolving black holes. New features in GISR are, the trends have a late time non-monotonic behaviour. This is because in its hamiltonian description \eqref{HamiltonianB}, the $u>v$ radiation and $u<v$ absorption terms are equally allowed. This means that the earlier radiated particles have  non-zero probability to come back and cause $c^{\vec{\omega}}_{u^n}(t)$'s Rabi-like oscillation. However, as $w$ increases, this late time oscillation will become negligible relative to the first cycle of increase-decrease variation. At the same time, the size of the matrix form Schr\"odinger equation \eqref{SchrodingerEq} grows exponentially.

\section{\bfseries Information Missing Puzzle} In GISR and its full quantum mechanical description, the causes of potential information missing effects are almost transparent. Firstly, the Wigner-Wiesskopf approximation, also known as Markovian approximation, includes forgetting history effects. Forgetting history implies information missing naturally. Technically \cite{dfzeng2021}, this happens in obtaining \eqref{WignerWiesskopfC} when shifting $c^{\phi}_{w^1}$'s history out of the integration symbol, 
\begin{align}
c^{o{\scriptscriptstyle1}}_{u\;\!\!^n}(t)\approx-i\int_{0}^tg_{u\;\!\!^n\!w\;\!\!^1}^{u{+}\omega=w}c^\phi_{w^1}(t')dt'
\rule{5mm}{0mm}
\\
\approx-ig_{u\;\!\!^n\!w\;\!\!^1}c^\phi_{w^1}(t)\int_{0}^{t\rightarrow\infty}e^{-i[\omega-(w-u)]t'}dt'
\end{align}
This doing introduces non-hermitian to the process and is the direct cause of thermal spectrum.  Secondly or more importantly, in Hawking type arguments, a black hole's radiation evolution is considered a sequence of events like
\beq{}
\raisebox{-7mm}{\includegraphics[totalheight=12mm]{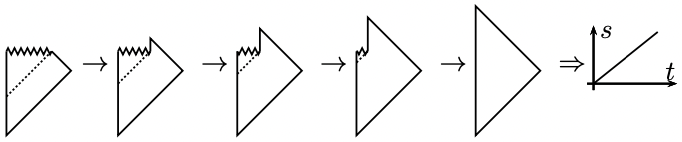}}
\label{SponRadiationChainA}
\eeq{}
While in the true quantum world, at any given time, the black hole size is not specifiable and the system can only be considered superposition of BR(black hole \& radiation)- combinationss of different mass-ratio,
\beq{}
\raisebox{-7mm}{\includegraphics[totalheight=11mm]{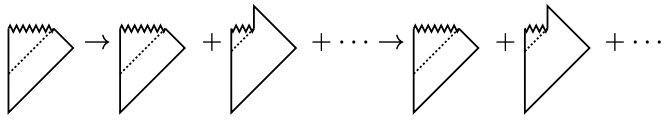}}
\label{SponRadiationChainC}
\eeq
where PC-diagram with different length of zigzag part denotes different size black holes and their corresponding radiation products.  Considerations \eqref{SponRadiationChainA} ignore entanglement superposition among BR-combinations of different mass-ratio, thus introducing entropies to the system artificially.

Mathematically, the process \eqref{SponRadiationChainC} can be written as
\begin{align}
|\psi_0\rangle\rightarrow|\psi_0\rangle{+}c_b^{t_1}|b\;\!\!^b\!r\;\!\!^b\rangle{+}c_m^{t_1}|b\;\!\!^m\!r\;\!\!^m\rangle{+}c_s^{t_1}|b\;\!\!^s\!r\;\!\!^s\rangle{+}|\psi_1\rangle
\label{stateSuperposition}
\\
\rightarrow|\psi_0\rangle{+}c_b^{t_2}|b\;\!\!^b\!r\;\!\!^b\rangle{+}c_m^{t_2}|b\;\!\!^m\!r\;\!\!^m\rangle{+}c_s^{t_2}|b\;\!\!^s\!r\;\!\!^s\rangle{+}|\psi_1\rangle
\rightarrow\cdots
\nonumber
\end{align}
where $|\psi_0\rangle$, $|\psi_1\rangle$ are the radiation-before, evaporation-after black hole state respectively,
$|{b\;\!\!^{b}r\;\!\!^{b}}\rangle$, $|{b\;\!\!^{m}r\;\!\!^{m}}\rangle$, $|{b\;\!\!^{s}r\;\!\!^{s}}\rangle$ are three typical but non-exhaustive intermediate state of big, median, small black holes with their radiation products. Tracing out the microscopic state of the black hole, we can write
\begin{align}
|\psi^t_{r}\rangle\sim\sqrt{\rho^0}\oplus c^t_b\sqrt{\rho^b}\oplus c^t_m\sqrt{\rho^m}\oplus c^t_s\sqrt{\rho^s}\oplus\sqrt{\rho^1}
\label{addMixedstate}
\\
\rho^b=\mathrm{tr}_{b^b}|b^br^b\rangle\langle{b^br^b}|,~\rho^m=\mathrm{tr}_{b^m}|b^mr^m\rangle\langle{b^mr^m}|
,~\cdots\nonumber
\end{align}
So the entropy of the radiation product can be calculated routinely $s=\mathrm{tr}_{rr'}\langle\psi^t_r|\psi^t_{r'}\rangle\ln\langle\psi^t_r|\psi^t_{r'}\rangle$, with result
\beq{}
s(t)=|c^t_b|^2s_{b\;\!\!^{b}r\;\!\!^{b}}+|c^t_m|^2s_{b\;\!\!^{m}r\;\!\!^{m}}+|c^t_s|^2s_{b\;\!\!^{s}r\;\!\!^{s}}
\Rightarrow
\raisebox{-2mm}{\includegraphics[totalheight=8mm]{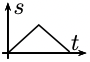}}
\label{entropySuperposition}
\eeq
where $s_{b\;\!\!^{b}r\;\!\!^{b}}$, $s_{b\;\!\!^{m}r\;\!\!^{m}}$, $s_{b\;\!\!^{s}r\;\!\!^{s}}$ denote entanglement entropies of the intermediate state black hole and their radiation products. At early times, $c_b^{t_1}\gg c^{t_1}_m\&c^{t_1}_s$,  so $s(t)$ is dominated by $s_{b\;\!\!^{b}r\;\!\!^{b}}$ and increases with time. As time passes by, $s(t)$ will be dominated by $s_{b\;\!\!^{m}r\;\!\!^{m}}$ and reach maximum on Page epoch, then decrease due to dominations of $s_{b\;\!\!^{s}r\;\!\!^{s}}$, and then Rabbi oscillate, as FIG.\ref{figWsixRadiationExample} displays.

From the viewpoint of GISR, the so called island formula or replica wormhole method \cite{ReplicaWormholeReview2006} are nothing but equivalent accounting for interferences between BR-combinations of different mass-ratio
\beq{}
\raisebox{-9mm}{\includegraphics[totalheight=18mm]{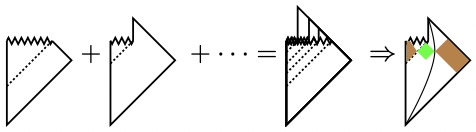}}
\eeq
This description yields Page-curve for Hawking radiation in 1+1 dimensional JT gravity, but it is not fundamental either on quantum mechanic or general relativity level \cite{mathur2021,giddings2022,stojkovic2020prd}. It cannot tell us where the ``missing'' information go. That is, when contributions from the replica wormhole channel is included, what should we measure to recover information stored in the initial black hole? In GISR's resolution, the answer to such questions are clear. The horizon size v.s. time relation of a black hole is determined by the initial state of the black hole through series of similarity factors appearing the hamiltonian of GISR and is measurable to retrieve initial state of the black hole directly, see FIG.\ref{figWsixRadiationExample} for concrete examples.

Essentially, what one mistakes in Hawking type argument for information missing is a kind of Schr\"odinger's cat misunderstanding. Binding one cat with each hawking particle will make this fact more easily understood. Since the hawking particles' emission is quantum mechanical,  the intermediate state of an evaporating black hole can only be considered superposition of many cat-groups with different number of died ones.
\beq{}
\raisebox{-6mm}{\!\!\!\includegraphics[totalheight=13mm]{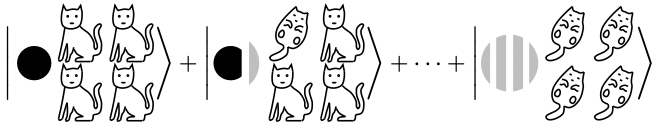}\!\!\!}
\eeq
But viewing the intermediate state as many different mass semi-classic black holes and their corresponding hawking particles, one implicitly measures the state of the cat-group
\beq{}
\raisebox{-6mm}{\!\!\!\includegraphics[totalheight=13mm]{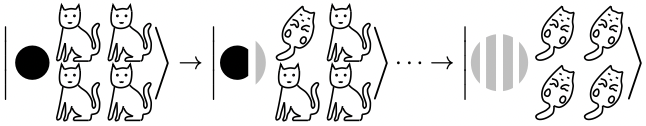}\!\!\!}
\eeq
Each of this measurement gives up knowledges on the quantum state of the hawking particles, especially the time interval between two successive cats' death, which is determined by similarity factors connecting the serial of initial-final state wave function of the black hole before and after each hawking particle's emission.  Neglecting this information will turn the beginning pure state to ending mixed one or beginning mixed state to ending more highly-mixed ones and produces entropy
\beq{}
{\Delta}S=(\frac{A_i}{4G}-\frac{A_{i+1}}{4G})-0
\eeq 
where $0$ denotes the entropy of initial pure state, $A_i$ and $A_{i+1}$ are the horizon area of the black hole on $i$ and $i{+}1$th measurement. Their difference quantise entropy of hawking particles emitted between this two measurements. The positivity of ${\Delta}S$ will cause the entropy of hawking particles increase monotonically. 

\section{Inside black holes} 

The idea of GISR dates back to 1970s \cite{Bekenstein1974a,Mukhanov1986,Bekenstein1997b}, during which Mukhanov and Bekenstein speculated atomic physics like interpretation for hawking radiation. However, due to their simple quantisation of black hole masses basing on adiabatic invariance of the horizon area, MB derived out discrete line shape spectrum for the radiation, which contradicts Hawking's thermal and continuous spectrum obviously. However, just as was shown in \cite{dfzeng2018a,dfzeng2018b,stojkovic2020unp,dfzeng2021,dfzeng2017,dfzeng2020}, the microscopic state of black holes can be quantised in such a way that their mass spectrum is continuous. So contradictions plaguing MB are avoidable. To see this directly, we start with the following exact black hole solutions family to the Einstein equation with nontrivial inside horizon structure
\begin{align}
ds^2_\mathrm{in}{=}{-}d\tau^2{+}\frac{\big[1{-}\big(\frac{2GM}{\varrho^3}\big)\!^\frac{1}{2}\!\frac{M'\!\varrho}{2M}\tau\big]^2\!d\varrho^2}{a[\tau,\varrho]}{+}a[\tau,\varrho]^2\varrho^2d\Omega^2_2
\label{genOSmetric}
\\
a[\tau,\varrho]\!=\!\big[1\!-\!\frac{3}{2}\big(\frac{2GM[\varrho]}{\varrho^3}\big)\!^\frac{1}{2}\tau\big]^{\!\frac{2}{3}}
,~M[\varrho\!\geqslant\!\varrho_\mathrm{max}]{=}M_\mathrm{tot}
\label{genOSscalefactor}
\\
a[\tau\!\in\!|_{\frac{p\;\!\!^\varrho}{4}}^{\frac{p\;\!\!^\varrho}{2}},\varrho]{=}{-}a[\frac{p\;\!\!^\varrho}{2}{-}\tau,\varrho]
,a[\tau|_{\frac{p\;\!\!^\varrho}{2}}^\frac{p\;\!\!^\varrho}{1},\varrho]{=}{-}a[p^\varrho{-}\tau,\varrho]
\\
a[\tau,\varrho]=a[\tau{+}p^{\varrho},\varrho]
,p^\varrho\equiv\frac{8}{3}\big(\frac{\varrho^3}{2GM[\varrho]}\big)\!^\frac{1}{2}
\rule{15mm}{0mm}
\label{aperiodic}
\end{align}
This describes a dust cluster's Over-Cross-Oscillation inside the horizon instead of aggregating on the central point and forming eternal singularities there. The energy-momentum tensor seeds the metric \eqref{genOSmetric}-\eqref{aperiodic} has forms  $T_{\mu\nu}\!=\!\mathrm{diagonal}\{\rho,0,0,0\}$ with $\rho\!=\!\frac{M'[\varrho]/8\pi\varrho^2}{a^{\!\frac{3}{2}}
\!\!+\!\!\frac{3GM'\tau^2}{4\varrho^2}
\!-\!\big(\frac{GM}{\varrho^3}\big)\!^\frac{1}{2}\frac{M'\varrho\tau}{2M}}$. Although dust is chosen here as proxies for matters consisting our black holes, when gravitation dominates all other interaction, this is a precise enough approximation for general matter sources. Reasons for their OCO can be attributed to the forward scattering amplitude domination instead of any unknown quantum gravitation effects. 

\begin{figure}[h]\begin{center}
\includegraphics[totalheight=35mm]{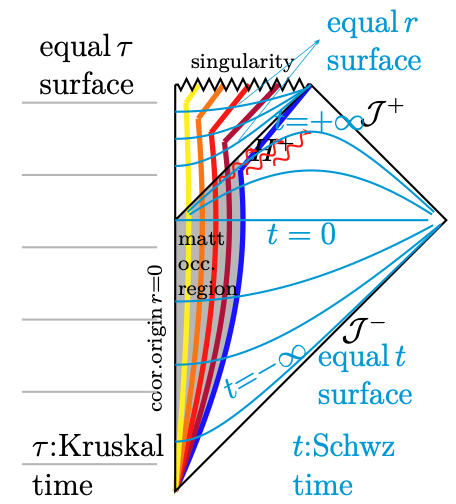}
\raisebox{3mm}{\includegraphics[totalheight=28mm]{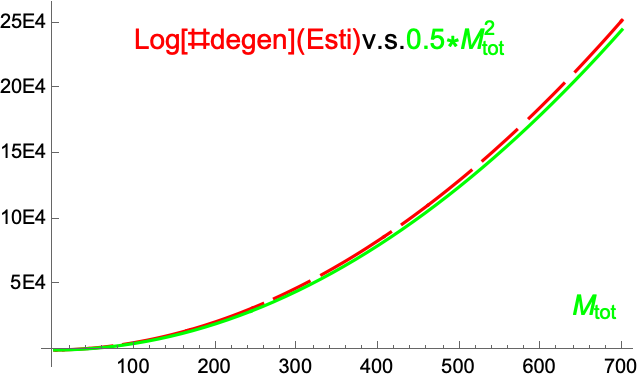}}
\includegraphics[totalheight=35mm]{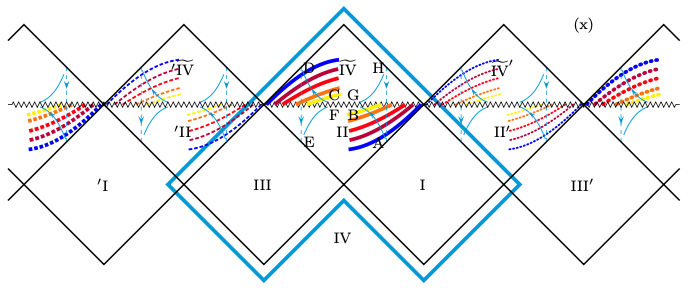}
\end{center}
\caption{In the outside observers' measurement, the formation of horizon and singularity due to gravitational collapse happens in the FINF, upper left. But in the parallel universe or ensemble spacetime implied by uncertainty principles, such events happen in their finite future. After quantisation,  the matter contents will exhibit $\exp(A/4G_N)$ ways of over cross oscillation inside the horizon determined by their total mass, upper right. Each happens in a parallel universe. Lower part is our extended PC-diagram in which the east and west semi-sphere of a full $R^3$ appear on the diagram independently, and the white hole region $\scriptstyle\mathrm{IV}$ is shifted from the past to future and becomes $\scriptstyle\widetilde{\mathrm{IV}}$. The parallel universe arises through general coordinate transformation like $r\rightarrow r^{in\pi}$. On this PC-diagram, a representation volume point of the outmost matter shell moves along traces like A-B-C-D-E-F-G-H-A$\cdots$, C is the antipodal of B, E is identified with D, and et al. }
\label{figPCdiagramSchwz}
\end{figure}

It is very important to note that OCO is a finite periodic oscillation only when measured with proper time defined on the black hole matter contents themselves. So in the above formulas, $\tau$, $\varrho$ and $\Omega_2$ are proper time and co-moving coordinate of the matter contents' volume element, $a[\tau,\varrho]$ the scale factor, $M[\varrho]$ the co-moving radial mass profile of the cluster at some arbitrary initial time whose concrete form is arbitrary on classic levels. To those outside horizon and coordinate time used observers, the black hole matter contents' oscillation happens in the Future of Infinite Future (FINF), see FIG.\ref{figPCdiagramSchwz}, upper left part for illustration. In reference \cite{dfzeng2021}, we resort uncertainty principle to allow this oscillation happen in finite coordinate time. This is essentially a parallel universe or ensemble spacetime interpretation for uncertainty principle. So to those observers, the oscillation of $a(\tau)$ and multiple possibility of $M[\varrho]$ is meaningful only in the sense of ensemble spacetime, or parallel universe implied by uncertainty principle. 

 Outside the matter occupation region our metric joins to the usual Schwarzschild black hole in Lema\"itre coordinate smoothly
\beq{}
ds^2_\mathrm{out}=-d\tau^2{+}\frac{d\varrho^2}{(1{-}\frac{3\tau}{2r_h})^\frac{2}{3}}{+}(1{-}\frac{3\tau}{2r_h})^\frac{4}{3}r_h^2d\Omega_2^2
\eeq
with $r_h=\frac{3}{8}p^{\varrho_\mathrm{max}}_\mathrm{eriod}=2GM_\mathrm{tot}$. This means that our inside horizon metric \eqref{genOSmetric}-\eqref{aperiodic} satisfies requirements of the no-hair theorem.
By shifting the white hole region to the future of black hole region and plotting the east\&west semi-sphere separately, we revise the usual Penrose Carter diagram and plot our black hole with OCO matter cores and parallel universe extension in FIG. \ref{figPCdiagramSchwz}, lower part. Our revision of Penrose-Carter diagram with east and west semi-sphere displayed independently on the map contain ideas similar with 't Hooft's antipodal identification \cite{tHooft2021}. From the figure, we easily see that our solutions respect the singularity theorem very well, i.e. all matters consisting of or falling into the black hole will reach the singularity in finite proper time \cite{Penrose1965,Hawking1976,geroch1979}. But such singularity is a periodic phenomena only instead an eternal one and end of physics.

Quantisation of the OCO structure \eqref{genOSmetric}-\eqref{aperiodic} can be done canonically.  Looking the whole matter core as a direct sum of many concentric shells, each moves freely under gravitations due to itself and more inside partners. 
\begin{align}
\Bigg\{\begin{array}{l}
h_i\dot{t}^2-h^{-1}_i\dot{r}^2=1
,~h_i=1-\frac{2GM_i}{r}
\\
\ddot{t}+\Gamma^{(i)t}_{tr}\dot{t}\dot{r}=0
\Rightarrow h_i\dot{t}=\gamma_i=\mathrm{const}
\end{array}
\\
\Rightarrow\dot{r}^2-\gamma^2_i+h_i=0,~\gamma_i^2\leqslant0
\rule{13mm}{0mm}
\label{eomShellClassic}
\end{align}
where $h_i$ is the function appearing in the effective geometry felt by the $i$-th shell, $ds^2=-h_idt^2+h_i^{-1}dr^2+r^2d\Omega^2$,  $\Gamma_i$ is the corresponding Christoffel symbol. For each shell $i$, we quantise equation \eqref{eomShellClassic} by looking it as an operator version hamiltonian constraint and introduce a wave function $\psi_i(r)$ to denote probability amplitude the shell be found of $r$-size, so that
\beq{}
\big[{-}\frac{\hbar^2}{2m_i}\partial_r^2{-}\frac{GM_im_i}{r}{-}\frac{\gamma^2_i\!-\!1}{2}m_i\big]\psi_i(r){=}0,~0{\leqslant}r{<}\infty
\eeq
where $M_i$ is the mass of the $i$-th shell and its inner partners, while $m_i$ that of the $i$-th shell only. Square integrability of $\psi_i$ requires that
\begin{align}
\psi_i=N_ie^{-x}xL_{n_i-1}^1(2x),x\equiv mr(1\mm\gamma_i^2)^\frac{1}{2}/\hbar
\\
n_i=\frac{GM_im_i}{\hbar(1\mm\gamma_i^2)^\frac{1}{2}}=0,1,2,\cdots, \gamma_i^2\leqslant0
\rule{10mm}{0mm}\label{enQuantizConditionSchwz}
\end{align}
where $L_{n_i-1}^1(2x)$ is the associated Lagurre polynomial and $N_i$ its normalisation in standard mathematics. We then direct-product all $\psi_i$s to get the wave functional of the whole matter core as follows
\beq{}
\Psi[M(r)]=\psi_0\otimes\psi_1\otimes\psi_2\cdots, 
\sum_im_i=M_\mathrm{tot}
\label{directProductWaveFunction}
\eeq
We will call objects described by this wave functional as OCO fuzzy ball, which is a pictorially similar but mathematically more precise(in the sense we can write out its wave function explicitly) fuzzy ball than string theory fuzzy balls. The quantisation condition \eqref{enQuantizConditionSchwz} and the sum rule in \eqref{directProductWaveFunction} do not require discreteness of the black hole mass spectrum. However, they indeed form complete constraints on how the matter core of a black hole can be considered big number of concentric shells and from what initial radial position each shell is released to freely OCO inside the horizon, i.e. degeneracies of the wave function\eqref{directProductWaveFunction}. We provide in FIG.\ref{figPCdiagramSchwz}, upper right part evidences that, the number of such degeneracies is consistent with the area law formula of Bekenstein-Hawking entropy, $Log[\#degen]\approx\frac{M_\mathrm{tot}^2}{2}\propto{Area}$, see \cite{dfzeng2018a,dfzeng2018b, dfzeng2020, dfzeng2017} for more calculation details and \cite{dfzeng2021} for a schematic proof road. 

Obviously by replacing the conventional Schwarzschild singularity with their matter contents' OCO motion, we assign black hole an atomic like inner structure thus allowing their atomic like radiation practicable. Just as we pointed out previously, this replacement does not break the singularity theorem. Instead, the classic OCO motion extends connotations of the singularity theorem; while the quantum OCO not only resolves the Schwarzschild singularity but also yields proper area law formula for Bekenstein-Hawking entropy. In this sense, we claim that OCO or OCO fuzzball a discovered structure instead of introduced picture for black hole microscopics. We expect it to appear also in a full quantum gravitational description of black holes.  It allows us to understand black hole physics the same way as we do with other objects, instead of customly building a new set as was done in the replica wormhole or quantum extremal surface \cite{ReplicaWormholeReview2006}.

\section{\bfseries Conclusion} We proposed in this work GISR as a new and universal mechanism for hawking radiation and a simple resolution to the black hole information missing puzzle. We provide an explicitly hermitian hamiltonian description for this mechanism and the microscopic structure of black holes supporting it which we call OCO fuzzy balls. Our hamiltonian description indicates that hawking particles carry information away from the black hole through changing its inner structures hence its evaporation progression. Viewing an evaporating black hole as series of time-dependent semi-classic objects, the conventional arguments for information missing ignore quantum interferences between BR-combinations of different mass-ratio thus introducing entropies to the system artificially and cause information missing puzzle. OCO fuzzy ball has a similar physical picture to string theory fuzzy balls but exactly spelled-out wave function and countable degeneracies consistent with the Bekenstein-Hawking entropy formula. OCO is a finite period oscillation only when measured with proper time associated with the volume element of the black hole matter contents. When measured with the outside horizon observer's coordinate time, this oscillation happens in the FINF, or parallel universe or ensemble spacetime required by the statistic interpretation of uncertainty principle. Both GISR and OCO fuzzy ball resort no physics beyond the standard general relativity and quantum mechanics. This means that, potentials of the canonic quantum mechanic approach to the theory of quantum gravity is worth of digging up further while new ideas such as replica wormholes on the basic feature of quantum gravitation need be examined more carefully. 

As discussion, we first note that GISR is a universal radiation which happens to all kinds of compositional objects. So it is very interesting to explore features or find evidences for such a radiation in, e.g. the direct S-wave to S-wave transition of usual atoms. Secondly, tidal effects of the OCO structured black holes are also observable from gravitational waves following from such objects' binary body merger \cite{cardoso2016,cardoso2017}. Thirdly, the OCO fuzzy ball interpretation of BH entropy is generalisable to higher dimensions and will give definite predictions \cite{dfzeng2018b,dfzeng2021}  for some real-number partition questions defined by constraints like \eqref{enQuantizConditionSchwz} and \eqref{directProductWaveFunction}. In one word, to accept that behind the information missing puzzle happened GISR is a big deal of small capital. Otherwise we need to answer, how to distinguish particles found around the black hole are due to Hawking Radiation or GISR? If we refute GISR from the beginning, anther question arises, why gravitation couplings between black holes and those to-be radiated particles do not cause radiation while electromagnetic couplings between the charged particles and photons do? See the following appendix for more reasonings on the rationality of GISR's universality.

\appendix

\section{\bfseries About the Universality of GISR}        
 
Many people think that hawking radiation is a privilege of black hole like objects with horizon. However, the vacuum definition and calculation of quantum field theory in curved spacetime \cite{BirrelDaviesTextbook,Carroll,Jacobson1996,Traschen} indicate that the prerequisite for hawking radiation is a future or past horizon instead of both. Almost all compositional objects with masses can have future horizon which can be used as the boundary of vacuum definition and cause hawking radiation. It is just basing on this fact that we conjecture GISR happens to all compositional objects with inner structures. To understand the logics behind this conjecture well, let us go through the relevant calculation and reasonings step by step.

i) A uniform accelerating observer in Minkowski spacetime can see thermal particles from the vacuum state defined by static observers. This is the so called Unruh effects.

ii) Hawking radiation can be looked as an Unruh effect measured by static observers sitting on fixed radial positions in a Schwarzshild (just an example) black hole outside, such observers are accelerating otherwise will fall towards the black hole due to gravitations.

iii) The definition of particles or vacuum is related with the global feature of spacetime  the observer is living in and fields he/she is examining. Three  types of vacuums are known explicitly, Hartle-Hawking, Unruh  and Boulware ones. All are defined through regularity conditions in Hadamard sense. That is, the two point correlation functions have asymptotics $G(x_1,x_2)\rightarrow\frac{U(x_1,x_2)}{4\pi^2\sigma_\epsilon}+V(x_1,x_2)\ln\sigma_\epsilon+W(x_1,x_2)$ as $|x_1{-}x_2|\equiv{}dx_{12}^2+2i\epsilon(t_1-t_2)+\epsilon^2\rightarrow0$, where $U$, $V$ and $W$ are regular functions.

\begin{figure}[h]
\bec{}\includegraphics[totalheight=50mm]{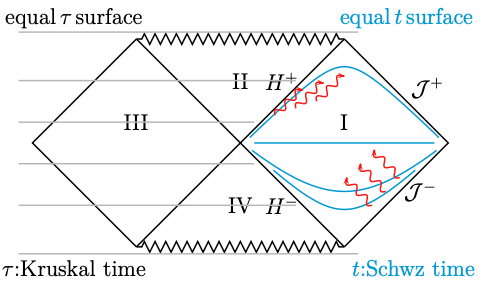}\eec
\caption{Figures for Hartle-Hawking vacuum definition}
\end{figure}         
iv) Hartle-Hawking vacuum is defined on eternal black hole background. It is regular on the past and future horizon $H^{\pm}$ at $r=2GM$, also on the past and future null infinity $\mathcal{J}^\pm$. The static observers in region I can see thermal radiations emitting from the black hole and equal flux of radiations from past null infinity. So the black hole is in thermal equilibrium with its environment, thus the nomenclature of eternal. At the the same time of radiate they also absorb.

\begin{figure}[h]
\bec{}\includegraphics[totalheight=50mm]{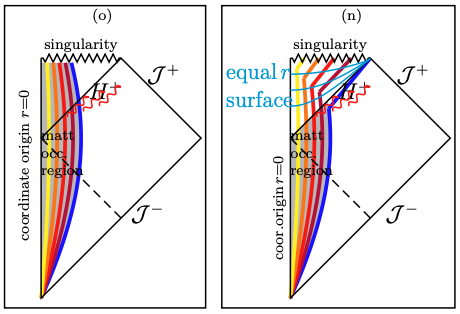}\eec
\caption{Figures for Unruh vacuum definition}
\label{figUnruhVacuum}
\end{figure}
v) Unruh vacuum is singular on the $H^-$ but regular in $H^+$. For this reason it is considered more appropriate for black holes forms through gravitational collapse. In such black holes $H^-$ is just an imagined surface instead of a happend one, see the dashed lines in Fig.\ref{figUnruhVacuum}. The static observers sitting outside the horizon can see thermal radiations from $H^+$ but not from $\mathcal{J}^-$. However, there is a key point worthy of noticing here. That is, the equal $r$-surfaces inside the horizon are more horizontal than vertical like those outside the horizon, see FIG.\ref{figUnruhVacuum} for illustrations. For this reason, the inside horizon part of the matter occupation region should be plotted as the right panel instead of the left one in the figure.

vi) Boulware vacuum is such a state that static observers sitting outside the horizon will not see particles escaping away or falling towards $\mathcal{J}^{\pm}$, thus they will not see hawking radiation. However, such a vacuum is singular on both $H^+$ and $H^-$. Its existence tells us that, regularity conditions on the horizon play key roles for our definition of vacuum state and thermal radiation flux.

vii) For reason vi), many people think that objects without horizon, such as neutron stars will not emit hawking radiation. However, this thought neglects an important fact that, in real black holes forms through gravitational collapse, the horizon is such a surface appears only in the infinite future of observers sitting 
\begin{figure}[h]
\bec{}\includegraphics[totalheight=50mm]{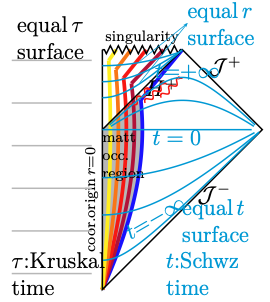}\eec
\caption{The formation of horizon through gravitational collapses needs infinite long Schwarzschild time. Hawking radiation happens during the collapsing process instead of the post horizon-formation era.}
\label{figFutureHorizon}
\end{figure}
outside the collapsing star. See Fig.\ref{figFutureHorizon} for references. Note that $H^+$ corresponds to $t=+\infty$ surface.

viii) General compositional objects with inner structures can always be considered a black hole to be, we can use their future event horizon $H^+$ as the place to impose regular conditions for vacuum state definition and get thermal radiation fluxes for observers sitting outside the matter occupation region.

ix) A possible argument against the previous reasoning may be, for objects without horizon, we have a time-like Killing vector throughout the spacetime. So we can use it to define positive frequency modes everywhere and match the Minkiowski modes at infinity. The resulting vacuum state would look like exactly of Boulware type, free of fluxes both at $\mathcal{J}^+$ and $\mathcal{J}^-$. The fact that Boulware vacuum is singular on the horizon will not bring us any trouble, since in this case, we have no horizon at all.

x) Answers to the counterview ix) are as follows, although we have a time like Killing vector through out the spacetime, its joint condition across the interface of matter occupation region happens in such a way that positive-frequency modes inside the matter region mixes positive and negative frequency ones outside it and vice versa. References \cite{Stojkovic2007,Stojkovic2008a,Stojkovic2008b,Stojkovic2009,Stojkovic2013,Stojkovic2014,Stojkovic2015} provides a big amount of  calculations and arguments for this fact. According to these works, hawking radiation happens during the collapsing process instead of the post horizon formation era or the ``black holes'' radiate away before they forms definitely.

Basing on the above analysis and reasoning, we think it may be rationale to assume that all compositional objects with inner structures can emit GISR. Nevertheless, we do not claim that GISR equals hawking radiation completely.

\section*{Acknowledgements}
This work is supported by NSFC grant no. {11875082}. 
The last stage of this work is finished at Niels Bohr Institute. We thank very much for warm hosts provided by Vitor Cardoso and support from Villum Investigator program supported by the VILLUM Foundation (grant no. VIL37766) and the DNRF Chair program (grant no. DNRF162) by the Danish National Research Foundation.




\begin{thebibliography}{00}


\bibitem{hawking1975cmp}
S. Hawking,
``black hole explosions'', {\em Nature} {\bf248} (1974) 30;
``Particle creation by black holes'',
{\em Comm. Math. Phys.} {\bf43} (1975) 199.

\bibitem{AdScftReview99}
O. Aharony, S. Gubser, J. Madacena, H. Ooguri, Y. Oz,
``Large N Field Theories, String Theory and Gravity'',
{\em Phys. Rept.} {\bf323} (2000) 183.

\bibitem{EREPR} 
J. Maldacena, L. Susskind,
``Cool horizons for entangled black holes'',
{\em Fortschr. Phys.}  {\bf 1–31} (2013),\href{https://doi.org/10.1002/prop.201300020}{DOI 10.1002/prop.201300020},
\href{https://arxiv.org/abs/1306.0533}{arXiv:1306.0533}

\bibitem{ReplicaWormholeReview2006}
A.~Almheiri, T.~Hartman, J.~Maldacena, E.~Shaghoulian and A.~Tajdini,
``The entropy of Hawking radiation,''
\href{https://arxiv.org/abs/2006.06872}{arXiv:2006.06872}.

\bibitem{fuzzball2009mathur}
S. Mathur,
``The information paradox: A pedagogical introduction'',
{\it Class. Quant. Grav.} {\bf26} (2009) 224001,
\href{https://arxiv.org/abs/0909.1038}{eprint:0909.1038}

\bibitem{dfzeng2018a}
Ding-fang Zeng, 
``Schwarzschild Fuzzball and Explicitly Unitary Hawking Radiations'',
{\em Nucl. Phys.}{\bf B930} (2018) 533-544,
\href{http://arxiv.org/abs/arXiv:1802.00675}{arXiv: 1802.00675}.

\bibitem{dfzeng2018b}
Ding-fang Zeng, 
``Information missing puzzle, where is hawking's error?'',
{\em Nucl. Phys.}{\bf B941} (2018) 665,
\href{http://arxiv.org/abs/arXiv:1804.06726}{arXiv: 1804.06726}.

\bibitem{stojkovic2020unp}
D. Dai, D. Minic, D. Stojkovic,
``On black holes as macroscopic quantum objects'',
\href{https://arxiv.org/abs/2006.09202}{arXiv:2006.09202}.

\bibitem{WeinbergSUSYtxtbook}
S. Weinberg
``The Quantum Theory of Fields (Vol-III Supersymmetry)'' , 
{\it CAMBRIGE UNIVERSITY PRESS} {\bf ISBN:9781139632638}(2013).

\bibitem{fireworksAMPS2012}
A. Almheiri, D. Marolf and J. Polchinski et al,
``Black Holes: Complementarity or Firewalls?''
{\it JHEP} {\bf1302} (2013) 062,
\href{http://arxiv.org/abs/1207.3123}{arXiv:1207.3123}

\bibitem{fireworksAMPS2013}
A. Almheiri, D. Marolf and J. Polchinski et al, 
``An Apologia for Firewalls'', 
{\it JHEP} {\bf1309}(2013) 018,
\href{http://arxiv.org/abs/1304.6483}{arXiv:1304.6483}.

\bibitem{dfzeng2021}
Ding-fang Zeng, 
``Spontaneous Radiation of black holes'',
{\em Nucl.Phys.} {\bf B977} (2022) 115722,
\href{https://arxiv.org/abs/2112.12531}{arXiv: 2112.12531}

\bibitem{Papadodimas2013}
Papadodimas, K., Raju, S. 
``An infalling observer in AdS/CFT'',
{\it JHEP} {\bf212} (2013),
\href{https://arxiv.org/abs/1211.6767}{1211.6767}

\bibitem{Stephen1302}
S. Hsu,
``Macroscopic superpositions and black hole unitarity'',
\href{https://arxiv.org/abs/1302.0451}{1302.0451}

\bibitem{Stephen1308}
S. Hsu,
``Factorization of unitarity and black hole firewalls'',
\href{https://arxiv.org/abs/1308.5686}{1308.5686}

\bibitem{Stephen2022}
X. Calmet, S. Hsu,
``Quantum Hair in Electrodynamics and Gravity'',
{\it EPL} 139 (2022) {\bf4}, 49001,
\href{https://arxiv.org/abs/2209.12798}{2209.12798}

\bibitem{NVW2013a}
Y. Nomura, J Varela, S. J. Weinberg,
``Black Holes, Information, and Hilbert Space for Quantum Gravity'',
{\it Phys. Rev. }{\bf D87} (2013) 084050,
\href{https://doi.org/10.1103/PhysRevD.87.084050}{https://doi.org/10.1103/PhysRevD.87.084050}

\bibitem{NVW2013b}
Y. Nomura, J Varela, S. J. Weinberg,
``Complementarity Endures: No Firewall for an Infalling Observer'',
{\it JHEP} {\bf 03} (2013) 059,
\href{https://doi.org/10.1007/JHEP03(2013)059}{https://doi.org/10.1007/JHEP03(2013)059}

\bibitem{ScullyQOtextbook}
M. Scully, \& M. Zubairy,
``Quantum Optics'', section 6.3,
 Cambridge University Press, 
 ISBN-13,978-0524235959.

\bibitem{mathur2021}
B. Guo, M. Hughes, S. Mathur, M. Mehta,
``Contrasting the fuzzball and wormhole paradigms for black holes'',
{\it Turk. J. Phys.} {\bf 45} (2021) 6, 281-365,
\href{https://arxiv.org/abs/2111.05295}{eprint: 2111.05295}

\bibitem{giddings2022}
S. Giddings,
``The deepest problem: some perspectives on quantum gravity'',
\href{https://arxiv.org/abs/2202.08292}{arXiv:2202.08292}

\bibitem{stojkovic2020prd}
D. Dai, D. Minic, D. Stojkovic, and C. Fu,
``Testing the ER=EPR conjecture'',
{\it Phys. Rev.} {\bf D102} (2020), 066004;
\href{https://arxiv.org/abs/2002.08178}{arXiv:2002.08178}.

\bibitem{Bekenstein1974a}
J. Bekenstein,
``The quantum mass spectrum of the Kerr black hole'',
\href{https://link.springer.com/article/10.1007/BF02762768}{Lett. Nuovo Cimento 11, 467 (1974)}.

\bibitem{Mukhanov1986}
V. Mukhanov, ``Are black holes quantized?'',
\href{https://ui.adsabs.harvard.edu/abs/1986JETPL..44...63M/abstract}{JETP Letters 44, 63 (1986)};
V Mukhanov in Complexity, Entropy and the Physics of Information: SFI Studies in the Sciences of Complexity, Vol. III, ed. W H Zurek (Addison–Wesley, New York, 1990).

\bibitem{Bekenstein1997b}
J. Bekenstein,
``Quantum black holes as atoms'',
Talk given at conference ``8th Marcel Grossmann Meeting on Recent Developments in Theoretical and Experimental General Relativity, Gravitation and Relativistic Field Theories'', p.92-111,
\href{https://arxiv.org/abs/gr-qc/9710076}{ePrint: gr-qc/9710076}
 
 \bibitem{dfzeng2017}
Ding-fang Zeng, 
``Resolving the Schwarzschild singularity in both classic and quantum gravities'',
{\em Nucl. Phys.} {\bf B917} 178-192,
\href{http://arxiv.org/abs/arXiv:1606.06178}{arXiv: 1606.06178}.

\bibitem{dfzeng2020} 
Ding-fang Zeng, 
``Exact inner metric and microscopic state of AdS$_3$-Schwarzschild BHs'',
{\em Nucl. Phys.}{\bf B954} (2020) 115001,
\href{https://arxiv.org/abs/1812.06777}{arXiv: 1812.06777}.

\bibitem{tHooft2021}
Gerard't Hooft,
``The black hole Firewall Transformation and Realism in Quantum Mechanics'',
{\em Universe} {\bf 2021} (2021) 7, 298,
\href{https://arxiv.org/abs/2106.11152}{lectures on black hole}

\bibitem{Penrose1965}
R. Penrose, 
``Gravitational Collapse and Space-Time Singularities''. {\it Phys. Rev. Lett.} {\bf14 (3)}, 57(1965); 

\bibitem{Hawking1976}
S. Hawking,
``Breakdown of Predictability in Gravitational Collapse'',
{\em Phys. Rev. } {\bf D14, } 2460 (1976)

\bibitem{geroch1979}
R. Geroch and G. Horowitz, 
``Global structure of spacetimes, in General Relativity: An Einstein Centenary Survey'', 
{\bf pp. 212–293}, 1979.

\bibitem{cardoso2016}
V. Cardoso, S. Hopper, C F. B. Macedo and et al,
``Gravitational-wave signatures of exotic compact objects and of quantum corrections at the horizon scale'',
{\it Phys. Rev.} {\bf D94} (2016) no.8, 084031,
\href{https://arxiv.org/abs/1608.08637}{arXiv: 1608.08637}.

\bibitem{cardoso2017}
V. Cardoso, P. Pani,
``Tests for the existence of horizons through gravitational wave echoes'',
{\em Nature Astronomy} {\bf1} (2017) 586-591,
\href{https://arxiv.org/abs/1709.01525}{1709.01525}.

\bibitem{BirrelDaviesTextbook}
N. D. Birrell and P. C. Davies,
``Quantum Fields in Curved Spacetime'',
Cambridge University Press 1984, ISBN 0 521 23385 2.

\bibitem{Carroll}
S. A. Carroll,
``Spacetime and Geometry'',
Cambrigde University Press 2019,
\href{https://doi.org/10.1017/9781108770385}{https://doi.org/10.1017/9781108770385}.

\bibitem{Jacobson1996}
T.A Jacobson, ``Introductory lectures on black hole thermodynamics'',
Lectures at University of Utrecht,
\href{http://physics.umd.edu/grt/taj/776b/lectures.pdf}{http://physics.umd.edu/grt/taj/776b/lectures.pdf}.

\bibitem{Traschen}
J. Traschen,
``An introduction to black hole evaporation'',
\href{https://arxiv.org/abs/gr-qc/0010055}{arXiv:gr-qc/0010055}.

\bibitem{Stojkovic2007}
T. Vachaspati, D. Stojkovic, L. M. Krauss,
``Observation of Incipient Black Holes and the Information Loss Problem'',
{\em Phys. Rev. } {\bf D76} (2007) 024005
\href{https://arxiv.org/abs/gr-qc/0609024}{gr-qc/0609024}

\bibitem{Stojkovic2008a}
T. Vachaspati, D. Stojkovic;
``Quantum Radiation from Quantum Gravitational Collapse'',
{\em Phys. Lett.} {\bf B663} (2008) 107-110,
\href{https://arxiv.org/abs/gr-qc/0701096}{ePrint: gr-qc/0701096}.

\bibitem{Stojkovic2008b}
E. Greenwood, D. Stojkovic,
``Quantum gravitational collapse: non-singularity and non-locality''
{\em JHEP} {\bf0806} (2008) 042,
\href{https://arxiv.org/abs/0802.4087}{0802.4087}

\bibitem{Stojkovic2009}
J. Wang, E. Greenwood, D. Stojkovic,
``Schr\"odinger formalism, black hole horizons and singularity behavior'',
{\em Phys. Rev.} {\bf D80} (2009) 124027,
\href{https://arxiv.org/abs/0906.3250}{0906.3250}

\bibitem{Stojkovic2013}
J Hutchinson, D. Stojkovic,
``Icezones instead of firewalls: extended entanglement beyond the event horizon and unitary evaporation of a black hole'',
{\em Class. Quant. Grav.} {\bf33} (2016) no.13, 135006,
\href{https://arxiv.org/abs/1307.5861}{ePrint: 1307.5861}

\bibitem{Stojkovic2014}
A. Saini, D. Stojkovic,
``Non-local (but also non-singular) physics at the last stages of gravitational collapse'',
{\em Phys. Rev.} {\bf D89} (2014) 044003
\href{https://arxiv.org/abs/1401.6182}{1401.6182}

\bibitem{Stojkovic2015}
A. Saini, D. Stojkovic,
``Radiation from a collapsing object is manifestly unitary'',
{\it Phys. Rev. Lett.} {\bf114} (2015), 111301;
\href{https://128.84.21.199/abs/1503.01487v3}{arXiv: 1503.01487}.

\end{thebibliography}


\end{document}